\documentstyle[preprint,aps,eqsecnum]{revtex}
 
\input{psfig}
\def\be{\begin{equation}}
\def\ee{\end{equation}}
\def\bea{\begin{eqnarray}} 
\def\eea{\end{eqnarray}}
\def\line{\hbox to \hsize}    
\def\frac #1#2{{#1\over #2}}

\def \V{{\bf v}}

\def\vev #1{{\langle #1\rangle}}
\def\1{\mbox{\bf 1}} 

\begin{document}
\draft 

\title{
Iordanskii Force and the Gravitational 
Aharonov-Bohm effect for a Moving Vortex
}

\author{ MICHAEL STONE}
\address{University of Illinois, Department of Physics\\ 1110 W. Green St.\\
Urbana, IL 61801 USA\\E-mail: m-stone5@uiuc.edu}   

\maketitle

\begin{abstract}

I discuss the scattering of phonons by a vortex moving with respect
to  a superfluid condensate.  This allows us to test  the
compatibility of the scattering-theory derivation of the Iordanskii
force with the galilean invariance of the underlying fluid dynamics.
In order to obtain the correct result we must retain $O(v_s^2)$ terms
in the sound-wave equation, and this reinforces the interpretation,
due to Volovik, of the Iordanskii force as an analogue of the
gravitational Bohm-Aharonov effect.

\end{abstract}

\pacs{PACS numbers: 
   67.40.Vs, 
   47.37.+q 
  }

\section{Introduction}

The problem of computing the transverse force acting
on a vortex in  a superfluid has  recently engendered
a certain amount of  controversy. If the vortex moves at a velocity
$\V_L$ while the superfluid and normal components have asymptotic velocities
$\V_s$ and $\V_n$ respectively, then the  most general form of the
transverse force per unit length that is consistent with galilean invariance 
({\it i.e.\/} depends only on velocity differences) can be
written\cite{wexler1}
\be
{\bf F} =  A \kappa \hat{\bf z}\times (\V_L-\V_s)
+ B\kappa \hat{\bf z}\times (\V_L-\V_n).
\ee
Here $\kappa$ is the magnitude of the quantum of circulation about the
vortex,
\be
\kappa = \frac{h}{m},
\ee
with $m$  the mass of a helium atom, 
and $\hat{\bf z}$ a unit tangent to the vortex line
in the  direction of the circulation.
In
the absence of any normal component, elementary fluid mechanics shows
that the momentum flux into the vortex core is  
\be
{\bf F}_M =  \rho_{tot} \kappa \hat{\bf z}\times (\V_L-\V_s),
\ee
where $\rho_{tot}$ is the total mass density. This is the Magnus
force\cite{lamb}.
Once  a normal component is present, however, a variety of different 
expressions for the coefficients $A$ and $B$ have been given in the literature.

It is generally accepted that the coefficient  $A$ is $\rho_s$\cite{hallvinen,donnelly}. An
appealing   thermodynamic argument for this has recently been given
by Wexler\cite{wexler1}.  The controversy stems from the claim of
Wexler\cite{wexler1}, and Thouless {\it et al.\/}\cite{thouless-etal} that the
coefficient of $\V_L$ in ${\bf F}$, is  also equal to
$\rho_s$.  Since this coefficient is $A+B$, their claim, if true,
would force $B$ to be zero --- thus ruling out the existence of the
second term, the {\it Iordanskii force\/}, which is supposed to
originate in a left/right asymmetry in the 
scattering  of quasi-particles by the vortex line\cite{iordanskii}.
Sonin\cite{sonin1,sonin2}, on the other hand, has presented a
detailed review of the scattering of phonons by a vortex line at rest
with respect to the superflow.  His analysis (which has been
challenged by Wexler and Thouless\cite{wexler2,wexlerthouless}, but which I
believe to be correct) shows that the asymmetry arises from a
fluid dynamical analog\cite{berry} of the Bohm-Aharonov
effect\cite{bohm-aharonov}, and gives the coefficient of $-\V_n$ as
$\rho_n$. Thus $B=\rho_n$. Combining this
value of $B$ with the accepted  value for $A$ gives  the  transverse
part of the force per unit length as
\be
{\bf F} = \rho_s \kappa \hat{\bf z}\times (\V_L-\V_s)
+ \rho_n \kappa \hat{\bf z}\times (\V_L-\V_n).
\label{EQ:iordanskii_a}
\ee 
This is the most commonly accepted expression for the force. When it is written in this
form the first term is usually refered to as the superfluid Magnus force and the
second term as the Iordanskii force.

Since the total density is $\rho_{  tot}=\rho_s+\rho_n$, equation (\ref{EQ:iordanskii_a})
may equally well  be written
\be
{\bf F} = \rho_{  tot} \kappa \hat{\bf z}\times (\V_L-\V_s)
+ \rho_n \kappa \hat{\bf z}\times (\V_s-\V_n).
\label{EQ:iordanskii_b}
\ee 
The first part of (\ref{EQ:iordanskii_b}) is  the momentum transfer
to the vortex due to the  condensate motion (and possibly {\it
this\/} should be called the superfluid Magnus force), so the second
term must be the force on the vortex due to phonon scattering.  Part
of the phonon force is responsible for reducing the coefficient of
$\V_s$ from $\rho_{ tot}$ to $\rho_s$. Notice that this expression
for the phonon force does not depend on the  motion of the vortex
line relative to either component of the fluid.  Although Sonin, who
works in the frame $\V_L=0$, writes $\rho_n(\V_s-\V_n)$ in his
expression for the phonon force, his analysis of the scattering
process is restricted to the situation where $\V_s=0$, {\it i.e\/} to
the case where there is no relative motion between the vortex line
and the condensate. The $\V_s$ part of the force is inferred from the
thermodynamic and galilean invariance argument given above. A direct
demonstration that the phonon force is independent of the relative
velocity of the vortex and the condensate, and hence
that coefficient of $\V_s$ in the phonon force is indeed equal to
$\rho_n$,  would provide a useful consistency test of the
conventional  scattering-theory derivation of
(\ref{EQ:iordanskii_a}), because the galilean invariance that went
into deducing this coefficient is not manifest in the linearized
sound wave equation.  In this paper I will provide such a
demonstration. In obtaining the result  we will find it useful to
consider the analogy, first pointed out by Volovik\cite{volovikBA},
of phonon vortex scattering with the {\it gravitational}
Bohm-Aharonov effect where particles are scattered by  a spinning
cosmic string.

In the next section I will review the acoustic Bohm-Aharonov effect
and rederive Sonin's results for the phonon force in the case that
the vortex is at rest with respect to the condensate. In section
three I will  extend these results to the case in which the vortex
moves relative to the condensate, and relate the momentum given to
the phonons by the vortex to  the time delay of signals passing on
either side of a cosmic string. In the last section  I will discuss the
apparent conflict between our results and the claims of Thouless {\it
et al.\/}, and some possible resolutions.  

\section{The Acoustic Bohm-Aharonov Effect}

The  scattering of  phonons by a superfluid vortex
was first studied by Fetter\cite{fetter}. 
The wave equation  used by most recent authors
\cite{sonin1,sonin2,wexler2,wexlerthouless} is 
\be
\frac{\partial^2\phi}{\partial t^2} =c^2 \nabla^2\phi-
2 (\V\cdot \nabla)\frac{\partial\phi}{\partial t}.
\label{EQ:soninwexler}
\ee  
Here $\phi$ is the velocity potential, $\V=\V_v$ is the velocity field of the vortex 
\be
((v_v)_x, (v_v)_y)= \frac {\kappa}{2\pi}\left( \frac {-y}{x^2+y^2}, \frac
{x}{x^2+y^2}\right),
\label{EQ:v_v}
\ee
and 
$c$  the speed of sound.

When the sound field $\phi$ has harmonic time 
dependence, $\phi({\bf r},t)= e^{-i\omega t} \phi({\bf r})$, equation
(\ref{EQ:soninwexler}) 
becomes 
\be
-\omega^2\phi =c^2 \nabla^2\phi+
2i\omega (\V_v\cdot \nabla)\phi.
\label{EQ:swharmonic}
\ee
We will be  interested in effects only to first order in the
circulation $\kappa$, therefore it  is natural to add harmless $O(\V_v^2)$ terms  
to (\ref{EQ:swharmonic}) so that it becomes the Schr{\"o}dinger equation 
for  unit charge particles  minimally coupled to a gauge
field ${\bf A} = \frac{\omega}{c^2} \V_v$, {\it viz. \/}
\be
-\omega^2\phi = c^2 \left(\nabla +i\frac{\omega}{c^2}
\V_v\right)^2\phi.
\label{EQ:bohm-aharonov}
\ee
Notice that this rewriting requires $\nabla\cdot\V_v=0$.

Equation (\ref{EQ:bohm-aharonov}) describes the
Bohm-Aharonov interaction  of particles with a threadlike  tube of
magnetic flux
in the gauge where $\nabla\cdot {\bf A}=0$.  The total flux in the
tube is  
\be
\Phi= \oint {\bf A}\cdot d{\bf r}=\frac{\omega}{c^2} \oint \V_v\cdot d{\bf r} 
=\frac{\omega}{c^2}\kappa.
\ee
Here the integration contour surrounds the vortex, which we have taken
to be at the 
origin of our coordinate system.
We will use the symbol
$\alpha$ to denote the ratio of this flux to the  Dirac  flux
quantum, $\Phi_0= 2\pi$.

\begin{figure}
\psfig{figure=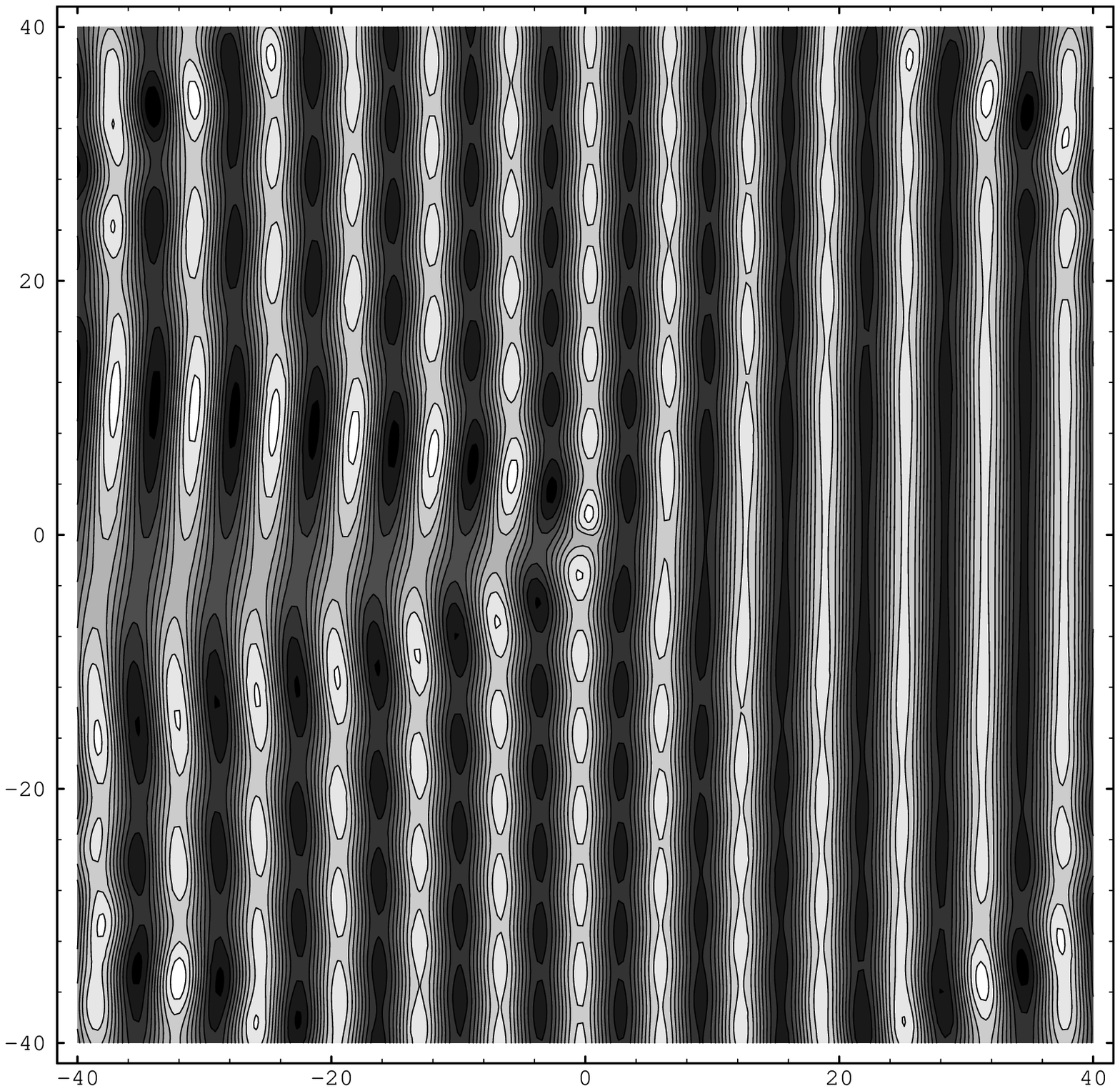,height=8.0in}
\caption{Bohm-Aharonov scattering for $\alpha=0.25$ \label{fig:bohm.25}}
\end{figure}

\begin{figure}
\psfig{figure=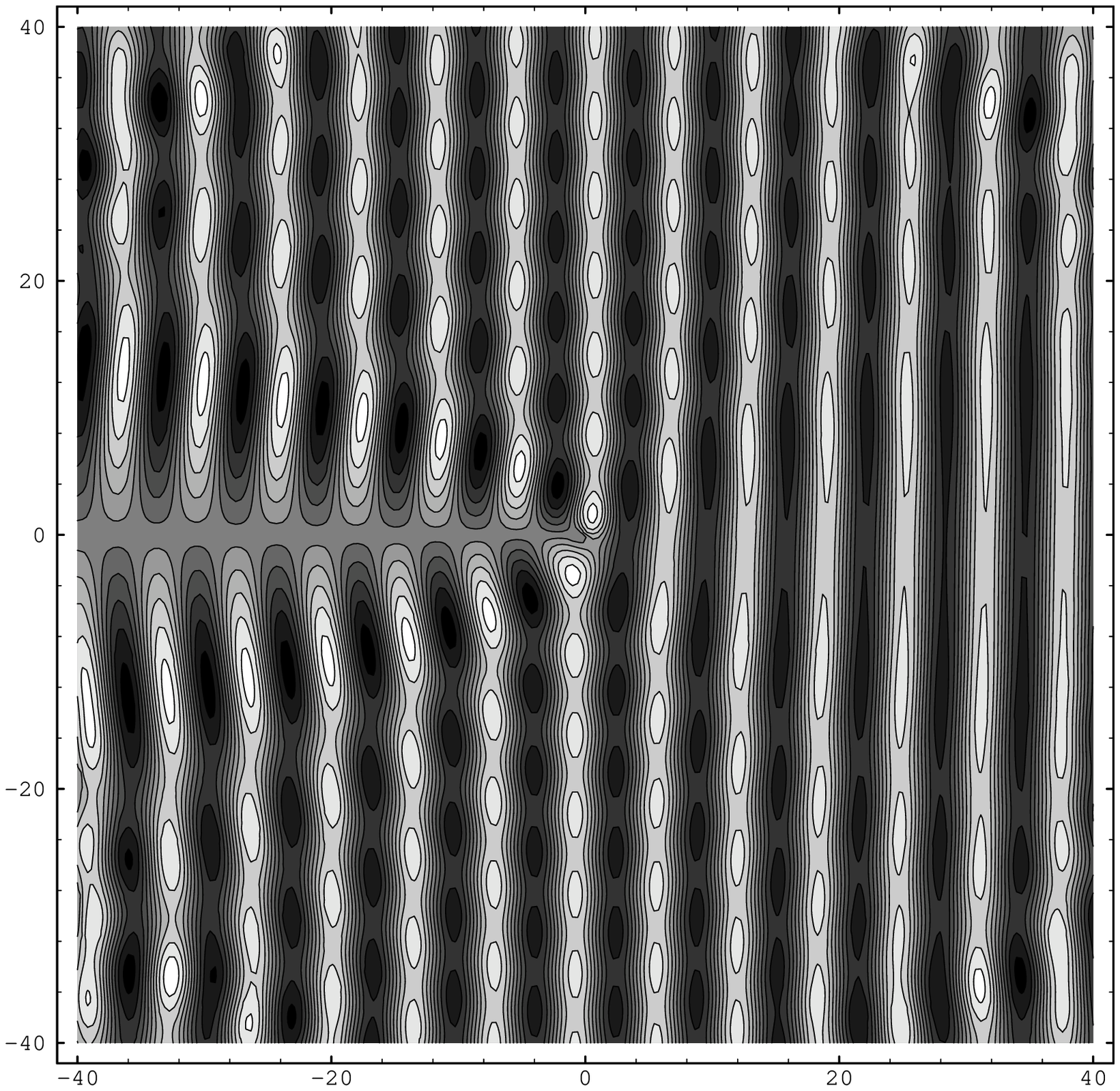,height=8.0in}
\caption{Bohm-Aharonov scattering for $\alpha=0.5$ \label{fig:bohm.5}}
\end{figure}

In their original paper\cite{bohm-aharonov}   Aharonov and  Bohm
provided a partial-wave series expansion for scattering of a plane
wave by the flux tube.  Figures \ref{fig:bohm.25} and
\ref{fig:bohm.5}   are numerical plots of   the sum of the first
forty terms in this expansion for the cases  where a plane wave
is incident from the right on flux tubes with $\alpha$ equal to
$0.25$ and $0.5$ respectively.  These plots should be compared to the
ripple tank photographs of surface waves interacting with a 
``bathtub'' vortex in \cite{berry}. The most noticable feature in
both the photographs and the plots is the ``seam'' or ``tear'' in the
wavefunction downstream of the flux tube. The incident plane wave is
cut in two by the flux tube and, apart from diffraction effects,
the upper and lower  halves of the incident wave propagate parallel
to each other  but with a relative phase shift of a fraction  $\alpha$ of a
wavelength.  It is in the region of the seam that the transverse
momentum imparted to the beam by the flux resides. Indeed in Fig
\ref{fig:bohm.25} one can plainly see that the wavefronts are
directed slightly upwards in this region.

The time-average momentum density in the sound wave is
$\vev{\rho_{(1)}\V_{(1)}}$
where $\V_{(1)}=\nabla\phi$ is the fluid velocity due to the sound wave and
\be
\rho_{(1)} = - \frac{\rho_{(0)}}{c^2}\{ \dot \phi + \V\cdot\nabla
\phi\}.
\ee
The angular brackets denote a time average.
(See the appendix for a derivation the expression for $\rho_{(1)}$.)
For a plane wave 
\be
\phi({\bf r}, t) ={\rm Re\,}\left\{\phi_0 e^{i{\bf k}\cdot {\bf r}-i\omega
t}\right\},
\ee
and with the background flow $\V$ vanishing, 
we have 
\be
\vev{\rho_{(1)}\V_{(1)}}= \frac 12 \rho_{(0)}
\frac{\omega}{c^2}|\phi_0|^2{\bf k}.
\ee
More generally  we find 
\be
\vev{\rho_{(1)}\V_{(1)}} = 
\frac 1{4i}  \frac {\rho_{(0)}\omega_r}{c^2}\left(\phi^*\nabla\phi-
(\nabla\phi^*)\phi\right),
\label{EQ:omega_r}
\ee
where $\omega_r$ is the frequency of the wave relative to  the 
fluid. (Notice that (\ref{EQ:omega_r}) is {\it not\/} the
``gauge invariant'' form of the  current for our minimally coupled
Schr{\"o}dinger equation.)

Once we are out of the region where $\V$ is significant, we can write
the momentum density as 
\be
\vev{\rho_{(1)}\V_{(1)}}= \frac 12 \rho_{(0)}
\frac{\omega}{c^2}|\phi_0|^2\nabla \chi,
\ee
where $\phi(x)=|\phi_0| e^{i\chi({\bf r})}$.  If we temporarily
ignore the reduction in the amplitude of the sound wave in the seam
region, we can find the total transverse momentum by integrating the
$y$ component of this momentum 
from one side of the seam to the other along a line parallel to the $y$ axis. 
The total transverse
momentum per unit length at abscissa $x$ is therefore
\be
\vev{p_y}=
\frac 12 \rho_{(0)}
\frac{\omega}{c^2}|\phi_0|^2 \Delta\chi(x),
\ee
where  $\Delta\chi(x)$, the phase difference across the seam, is zero long 
before the sound wave interacts with the  vortex, and  
\be
\Delta \chi(x) =2\pi \alpha = \frac{\omega}{c^2}\kappa
\ee
well after the sound waves have passed the vortex. 
In this manner the transverse momentum is  found by examining the
wave at large (but not infinite) impact parameter, and the result is
insensitive to details such as diffraction effects.

The transverse momentum per unit length of the seam can also be
written
\be
 \left(\frac 12 \rho_{(0)}
\frac{\omega}{c^2}|\phi_0|^2 k\right) \cdot \frac 1 k\cdot  \frac {\omega}{c^2}\kappa
= \vev{j_{\rm ph}}\cdot\frac 1 k \cdot \frac {\omega}{c^2}\kappa,
\ee
where $ \vev{j_{\rm ph}}$ is the mass current, or momentum density
in the unperturbed sound wave. If a finite pulse of sound is sent past 
the vortex then  a length of seam equal
to the group velocity of the waves (here $c$) is  created every
second. The  transverse force is  the rate of creation of
transverse momentum and this is 
\be
\vev{\dot P_\perp} =\vev{j_{\rm ph}}\cdot \frac 1 k \cdot \frac
{\omega}{c^2}\kappa\cdot c = \vev{j_{\rm ph}}\kappa
\label{EQ:statvforce}
\ee
in agreement with ref \cite{sonin2}.  

The mass flux  due to phonons in the two fluid model is
\be 
\vev{{\bf j}_{\rm ph}} =\rho_n(\V_n-\V_s),
\ee
and we can find the thermal average of the phonon force by substituting this in
 (\ref{EQ:statvforce}).   
Since we have so far assumed that $\V_s$ is zero  we have, however, only established
the $\rho_n\kappa \V_n$ part of the Iordanskii force.

A more rigorous approach to computing the momentum given to the
phonons exploits the momentum flux tensor $\Pi_{ij}^{\rm phon}$. The relevant terms are 
\be
\Pi_{ij}^{\rm phon}= \rho_{(0)}\vev{v_{(1)i}v_{(1)j}} + (v_v)_i\vev{\rho_{(1)}v_{(1)j}}
+(v_v)_j\vev{\rho_{(1)}v_{(1)i}}.
\ee
The only terms contributing to
$\Pi_{xy}^{\rm phon}=\Pi_{yx}^{\rm phon}$ to first order in $\V_v$  turn out to be
\be
\Pi_{xy}^{\rm phon} = \left(\Pi_{xy}^{\rm phon}\right)_{\V_v=0}+\frac ck\left(\partial_y \chi +\frac kc
(v_v)_y\right)\vev{(j_{\rm ph})_x},
\ee 
while $\Pi_{yy}^{\rm phon}$ is of at least second order in $\V_v$ and can be
neglected.

For most of the $x,y$ plane we may use the eikonal approximation for the phase
$\chi$,
\be
\chi(x,y)= kx - \frac {\omega}{c^2} \int_{-\infty}^{(x,y)} (v_v)_x
\,dx'.
\ee
Here the integral is taken along the line from the point $(-\infty,y)$ to
$(x,y)$.
(The eikonal approximation  will be described further in the next section.) 
From  the formulae for $(v_v)_x, (v_v)_y$ in (\ref{EQ:v_v}) we find  
\be
\int_{-\infty}^x (v_v)_x\, dx = -\frac {\kappa}{2\pi} \frac {y}{|y|}\left(\tan^{-1} \frac {x}{|y|}+\frac \pi 2\right). 
\ee
This expression is continuous across the negative $x$ axis, but  
jumps discontinuously by $\kappa$ across the positive $x$
axis.
We therefore find that the eikonal phase  has the expected  $\omega\kappa/c^2$ discontinuity across 
the seam. The physical wave
fronts, of course, smoothly interpolate the phase across the seam. We
immediately see that outside this interpolating zone, in the region
where 
the eikonal approximation to the phase is valid, we have 
\be
 \partial_y \chi +\frac kc
(v_v)_y=0. 
\ee
This means that $\Pi_{xy}$ is zero outside  the seam region. Thus
the  flux of $p_y$ through any curve vanishes unless it intersects
the seam.  Indeed we find that the only regions that have a net
transverse momentum flux out of them are those that include the
vortex. For these the  momentum flux is entirely due to the
interpolating phase and comes out to be   $\vev{(j_{\rm
ph})_x}\kappa$ as found by the previous, more intuitive, method.

The transverse momentum flux tensor vanishes because the transverse component of  ${\bf k}$ 
acquired by  interaction with the flow is cancelled by the transverse component of $\V_v$  
when they are combined to form the group velocity
\be
\frac{\partial \omega}{\partial {\bf k}}=\frac ck {\bf k} +
\V_v.
\ee   
The classical phonon trajectories therefore remain straight, and, just 
as in the ordinary Bohm-Aharonov effect, 
the transverse momentum is consequence of the wave-particle duality. 

A more detailed analysis would take into account the reduction of the
amplitude of the sound wave in the seam region. It is well understood
from the theory of  Bohm-Aharovov scattering\cite{shelankov} that the effect of this
is to replace $\Phi$ by $\sin \Phi$ in the force
equation. (Inspection of Fig \ref{fig:bohm.5} where
$\Phi =\pi$ shows that that the sound wave amplitude is exactly  zero
in the seam).
Using $c=230\, {\rm m}{\rm s}^{-1}$ for the speed of sound in liquid helium, we find 
\be
\alpha= \frac{\omega}{c^2} \frac{\hbar}{m} = 0.035 E_{\rm phon}
\ee
where $E_{\rm phon}$ is the energy of the phonon measured in degrees
Kelvin. We see that 
$\Phi$ is small at  temperatures below $0.2\,$K where phonons dominate the scattering process, so the correction 
will be unimportant.

\section{Moving the Vortex}

So far we have merely reproduced the results of \cite{sonin2}.  We
now  extend our analysis to the case in which the vortex moves with
respect to the condensate. So as to retain a time independent
equation we will keep the vortex fixed at the origin, but allow a
non-zero asymptotic $\V_s$.   For simplicity of description we will
initially consider only the case where the uniform $\V_s$ is in the
direction of propagation of the sound wave, which as before we take
to be the $+x$ direction.  We will write $\V_s=U \hat{\bf x}$.
In this case the length of seam created per second is $c+U$. We  need
to find the phase shift between two halves of the wavefront to
complete the computation.

It is tempting to simply replace the $\V_v$ in
(\ref{EQ:bohm-aharonov}) by $\V=U \hat{\bf x}+\V_v$, but this will
not serve to give the correct answer. Because of the Doppler shift, the
frequency is now related to the wavelength by $\omega=(c+U)k$, so for
a wave with the same $k$ the frequency, and hence the effective flux
\be
\Phi= \frac{\omega}{c^2}\oint \V_v\cdot d{\bf r}
\ee
is increased, and this  makes the phase shift larger. This is not what we expect, and
is incorrect. The terms added to the sound-wave equation to
make it into the minimally coupled Schr{\"o}dinger equation are no longer
harmless. This is  because even when we work only to $O(v_v)$ accuracy, we must 
not neglect $O(U^2)$ terms.

We require a more accurate equation. In the appendix we show that
the relevent equation is that  given by Unruh  \cite{unruh1,unruh2}
\be
\left(\frac{\partial}{\partial t}+\nabla\cdot\V\right)\frac
{\rho}{c^2}\left(\frac{\partial}{\partial t}+\V\cdot\nabla\right)\phi=
\nabla(\rho\nabla\phi),
\ee
who interprets his equation as that of a scalar field propagating in a
curved space-time background.

We can find the phase offset by working at large impact parameter where $\phi$ is
essentially independent of $y$, and where also $\partial_x \V_v$ is
negligeable. We can therefore set
\be
\phi=\varphi(x)e^{ikx-i\omega t}
\ee
with  $\omega= (c+U)k$, and  expect $\varphi$ to be slowly varying.
The equation for $\phi$ becomes
\be
-\omega^2 \phi -2i\omega v_x \partial_x \phi +v_x^2 \partial^2_{xx}\phi=
c^2\partial^2_{xx}\phi
\label{EQ:onedimapprox}
\ee
where $v_x= (v_v)_x+U$.
Since $\varphi(x)$ is slowly varying we can ignore
$\partial^2_{xx}\varphi(x)$. Doing so we find that (\ref{EQ:onedimapprox}) reduces to 
\be
(v_v)_x k\varphi -i(U+c)\partial_x\varphi=0.
\ee
(I have ignored terms containing $v_v$ in the coefficient of $\partial_x\varphi$ 
because they will not affect our
result to $O(\kappa)$.) 
This gives
\be
\varphi(x)=e^{-i\chi(x)}
\ee
with
\be
\chi(x) = \frac k{U+c} \int^x_{-\infty} (v_v)_x dx'. 
\ee
We see that as $x\to +\infty$ the phase offset of the two wavefronts
becomes
\be
\Delta\chi = \frac k{U+c} \oint \V_v \cdot d{\bf r}= \frac k{U+c}
\kappa.
\ee
The factor of $U+c$ cancels against the length of seam being created  per second to give
\be
\vev{\dot P_\perp}=\vev{j_{\rm ph}}\kappa
\ee
as before. 

This result can be confirmed by examining the momentum flux tensor.
The  $O(\V_v)$ part of $\Pi_{xy}$ is now
\be
\Pi_{xy}= \left((c+U)\frac 1k \partial_y \chi +
(v_v)_y\right)\vev{(j_{\rm ph})_x},
\ee
where we  have included  a non-zero contribution  from
$U\vev{\rho_{(1)}v_{(1)y}}$.
Once again we see that outside the seam region the gradient of $\chi$ cancels the 
$(v_v)_y$ advective term, and that the discontinuity across the seam provides
the  momentum flux obtained in the previous paragraph.

The wavefront offset can also be calculated from  the time delay
between phonons passing on other side of the vortex.  Since the
phonon trajectories  with large impact parameters are hardly
deflected, we can find them as the null geodesics of the simplified
form of the Unruh metric (\ref{EQ:unruh_simplified})
\be
ds^2=\frac \rho c
\left\{-\left(dx-(v+c)dt\right)\left(dx-(v-c)dt\right)-dy^2-dz^2\right\}.
\ee
The null geodesics are given by
\be
\frac {dx}{dt}= v\pm c.
\ee
In our present case, the time of arrival of a signal at the point
$(x,y)$ is 
\be
t(x,y) = \int^{(x,y)}_{-\infty} \frac{dx}{U+c+(v_v)_x}=  const.- \frac 1{(U+c)^2} \int^{(x,y)}_{-\infty}  (v_v)_x dx
+ O(v_v^2).
\ee 
We convert the time delay to a phase shift by multiplying by $\omega=(U+c)k$. Again 
we find that the relative phase shift between waves that pass above and below the vortex to be
\be
\Delta \chi = \frac k{U+c} \kappa.
\ee 
As described by Volovik \cite{volovikBA} this   
separation in the time of arrival for  signals passing abitrarily far from the
vortex is characteristic of a ``spinning cosmic string''.

Matters  become slightly more complicated when the uniform background
superflow $\V_s$ 
is not oriented parallel (or anti-parallel) to the incident phonon flux. 
After a little work we find that the eikonal
equation becomes
\be
({\bf U}^g\cdot \nabla) \chi + {\bf k}\cdot \V_v=0,
\ee
where ${\bf U}^g = c \hat {\bf k}
+ \V_s$.
The phase discontinuity seam no longer lies exactly along the $x$
axis,  but instead  points in the direction of ${\bf U}^g$.
The rate of transverse momentum production
does depend on the angle this vector makes with the x axis, but the
effects are $O(U^2/c^2)$. They do not seem to be worth working  out in detail since 
at this order we should also include  compressibility 
effects and the dependence of
$\rho_n$ and $\rho_s$ on $(\V_n-\V_s)^2/c^2$.

\section{Discussion}

We have computed the  force on a vortex due to the scattering of
phonons in the case where the vortex is moving with respect to the
condensate. We have shown that to first order in $|\V_s-\V_L|/c$ the rate of transverse 
momentum production is independent of the relative velocity of 
the vortex and the condensate, 
and that the Bohm-Aharonov phase shift of the
phonon wavefront passing on either side of the vortex leads to a
force
\be  
{\bf F}_{ {\rm phon}}=\rho_n \kappa \hat{\bf z}\times
(\V_s-\V_n),
\ee
which is consistent with galilean invariance. This supports the view
that the Bohm-Aharonov analysis of the phonon force is correct.

The expression we have obtained for the phonon force leads to the
commonly accepted form of the Iordanskii force and so disagrees with
the claims of Thouless {\it et al.\/} that it vanishes identically.
Their claim is based on earlier work by Thouless, Ao and Niu
(TAN)\cite{TAN} which establishes a general theorem relating the
force on a vortex to the circulation of momentum at infinity.
Since there are no impurities present, this theorem should hold here
and we have a puzzle that needs to be resolved.

In response to criticism of their
claims\cite{hallhook,sonin_comment}, Thouless {\it et
al.}\cite{TANWcomment}  have given two possible explanations for  the
discrepancy. The first suggests that  the computation of the cross
section asymmetry is mathematically flawed because a conditionally
convergent series is mistreated.  If this were correct, their
objection would also apply to the case where $\V_L=\V_n$, but we have
shown here that the scattering asymmetry correctly predicts the
reduction of the coefficient of $\V_s$ from $\rho_{tot}$ to $\rho_s$.
This explanation seems unlikely therefore.

The second possible explanation is more subtle. In  the calculations 
presented here, and in all earlier
work on phonon scattering, the incident flux of phonons is calculated
by assuming a thermal phonon distribution derived from the asymptotic
$\V_n$ and $\V_s$, {\it i.e.\/} a distribution that does not seem to take
into account the effect of the local vortex flow field $\V_v$.  Because the
assumed phonon flux to the left and right of the vortex is the same,
it appears that the phonons make an equal reduction in the fluid momentum on either
side of the vortex. In other words they appear not to change the  value of the
total momentum circulation 
\be 
K_{tot} = \oint \vev{\rho \V}\cdot
d{\bf r},  
\ee 
so that it  remains $\rho_{tot}h/m$ instead of being reduced to $\rho_s h/m$. 
If this were correct there would be
circulation in the normal fluid as well as in the superfluid
component. If we include this unphysical circulation in the TAN formula, then
it agrees with the calculated
Iordanskii force.  

The problem with this explanantion is that the motion of the phonons can be
derived from a hamiltonian, $H= c|k|+{\bf v}_v\cdot {\bf k}$.
Consequently
Liouville's theorem applies to the distribution function. Even in the
absence of phonon-phonon interactions, a phonon distribution function
that is in thermal equilibrium will remain in equilibrium and so 
phonons arriving from infinity will have their local distribution
modified by the flow field so as to correctly bring about the expected
reduction in the circulation.  Although there are an
equal number number of phonons passing to the right or to the left of
the vortex, those moving against the superflow are going slower,
so their {\it number  density\/} is higher.  This means that the phonon
momentum opposing  the $\rho_{tot}\V_v$ {\it is\/} left-right
asymmetric as it should be.

There remains a contradiction, therefore. 

What are we to conclude?  The TAN arguments implicitly make use of
the change in the total momentum of the medium outside the vortex. In
fluid mechanics it  is well known that the total momentum associated
with a system of vortices is ill-defined, being given by a
conditionally convergent integral. This problem is usually dealt with
defining the {\it impulse\/}\cite{lamb} of the vortex system.  The
impulse  is defined in terms  the velocity potential in the vicinity
of the vortex system, and does not have contributions from the
effects of distant boundaries (the ultimate origin of the ill-defined
 momentum). Perhaps this is at the root of the problem. On the
other hand the present calculation  might be also described as
na{\"\i}ve. We are not distinguishing between the true newtonian
momentum and the {\it pseudo-momentum\/}\cite{peierls} possessed by the phonons.
However pseudo-momentum is usually exactly what is needed for
computing forces on immersed objects. Clearly,
more work is needed to resolve the paradox.

\section{Acknowledgements}

This work was supported by grant NSF-DMR-98-17941. I would like to
thank Edouard Sonin and Andrei Shelankov for useful e-mail discussions and also David
Thouless, Ping Ao and  {\v S}imon Kos for many useful conversations.

\appendix
\section{The Unruh Equation} 

The homentropic potential flow of   a 
fluid is derivable from the action
\be
S= \int d^4x\left\{ \rho \dot \phi +\frac 12 \rho (\nabla\phi)^2
+ V(\rho)\right\}.
\ee
Here $\rho$ is the mass density and $\nabla
\phi=\V$, the fluid velocity.
Varying $S$ with respect to $\phi$ gives the continuity equation
\be
\dot \rho + \nabla\cdot (\rho \nabla \phi)=0,
\label{EQ:continuity}
\ee
while varying with respect to $\rho$ gives Bernoulli's equation
\be
\dot\phi +\frac 12 (\nabla\phi)^2 + \mu(\rho)=0,
\label{EQ:bernouilli}
\ee
with $\mu(\rho) = d V/d \rho$. 

In order to consider the propagation of sound waves in the background
flow, set
\bea
\phi&=&\phi_{(0)} + \phi_{(1)}\nonumber\\
\rho&=&\rho_{(0)} + \rho_{(1)}+\cdots 
\eea
where $\phi_{(0)}$ and $\rho_{(0)}$ obey the equations of motion,  and  $\phi_{(1)}$ and $\rho_{(1)}$
are small amplitude perturbations. Expanding $S$ to quadratic order in
the perturbations gives
\be
S=S_0+ \int d^4x\left\{\rho_{(1)} \dot \phi_{(1)}
+ \frac 12 \left(\frac {c^2}{\rho_{(0)}}\right) \rho_{(1)}^2
 +\frac 12 \rho_{(0)} (\nabla\phi_{(1)})^2
+ \rho_{(1)}\V\cdot\nabla \phi_{(1)} \right\}. 
\ee
(The terms linear in the perturbations  vanish because of
our assumption that the zeroth order variables obey the equation of motion.)
Here $\V \equiv \V_{(0)}= \nabla\phi_{(0)}$.  The speed of sound, $c$, is defined by   
\be
\frac {c^2} {\rho_{(0)}}= \left.\frac {d \mu}{d \rho}\right|_{\rho_{(0)}}, 
\ee
or more familiarly
\be
c^2= \frac {d P}{d \rho}.
\ee

Since the new action is quadratic in $\rho_{(1)}$, we can eliminate it
via its equation of motion
\be
\rho_{(1)} = - \frac{\rho_{(0)}}{c^2}\{ \dot \phi_{(1)}+ \V\cdot\nabla
\phi_{(1)}\}.
\ee
We find the effective action for the sound waves in the background flow $\V$  to be 
\be
S_{(2)} = \int d^4x\left\{ \frac 12 \rho_{(0)} (\nabla\phi_{(1)})^2 - \frac {\rho_{(0)}} {2 c^2}
(\dot \phi_{(1)}+ \V\cdot\nabla \phi_{(1)})^2\right\}.
\ee
After changing an overall sign for convenience, we  can write this as 
\be 
S=  \int d^4x\,\frac 12  \sqrt{-g} g^{\mu\nu} \partial_\mu \phi_{(1)} \partial_\nu
\phi_{(1)},
\ee
where 
\be
\sqrt{-g}g^{\mu\nu} = \frac {\rho_{(0)}}{c^2}\left(\matrix{ 1, & \V^T \cr
                                                            \V,& \V\V^T - c^2{\bf
1}\cr}\right).
\label{EQ:unruh_metric_up}
\ee  
(We use the convention that greek letters run over all four space-time
indices with $0\equiv t$, while roman indices refer to the spatial components.)  

The resultant equation of motion
\be
\frac 1{\sqrt{-g}} \partial_\mu {\sqrt{-g}}g^{\mu\nu}\partial_\nu
\phi_{(1)}=0,
\ee
is
\be
\left(\frac{\partial}{\partial t}+\nabla\cdot\V\right)\frac
{\rho_{(0)}}{c^2}\left(\frac{\partial}{\partial t}+\V\cdot\nabla\right)\phi_{(1)}=
\nabla(\rho_{(0)}\nabla\phi_{(1)}).
\label{EQ:unruheq}
\ee
This equation and its interpretation as the wave equation for 
a scalar field propagating in the background space-time  metric 
(\ref{EQ:unruh_metric_up}) is due to
Unruh \cite{unruh1,unruh2}.

In four dimensions we have  $\sqrt{-g}= \rho_{(0)}^2/c$ and
\be
g_{\mu\nu}= \frac {\rho_{(0)}}{c}\left(\matrix{ c^2-v^2, & \V^T \cr
                                                            \V,& -{\bf
1}\cr}\right).
\label{EQ:unruh_metric_down}
\ee
The associated space-time interval is therefore
\be
ds^2= \frac {\rho_{(0)}} c \left\{c^2dt^2
-\delta_{ij}(dx^i-v^idt)(dx^j-v^jdt)\right\}.
\ee
Up to the overall  conformal factor $\frac {\rho_{(0)}} c$ we see
that $c$ and $-v^i$ play the role of the lapse function and shift
vector appearing in  the Arnowitt-Deser-Misner (ADM) formalism of general
relativity\cite{MTW}. A conformal factor does not affect  null
geodesics, and so variations in $\rho_{(0)}$ do not influence the ray tracing for the
sound waves.

It is also sometimes convenient to write 
\be
ds^2=\frac {\rho_{(0)}}{ c} \left\{(c^2-v^2)\left(dt+ \frac{v^i
dx^i}{c^2-v^2}\right)^2-
\left(\delta_{ij}+ \frac{v^iv^j}{c^2-v^2}\right)dx^idx^j\right\}.
\ee
When $\V$ is  in the $x$
direction only, we can also rewrite $ds^2$ as
\be
ds^2=\frac {\rho_{(0)}} {c} \left\{-\left(dx-(v+c)dt\right)\left(dx-(v-c)dt\right)
-dy^2-dz^2\right\}.
\label{EQ:unruh_simplified}
\ee
This shows  that the $x-t$ plane  null geodesics coincide with the expected
characteristics of the wave equation in the background flow.


\eject
\end{document}